\documentclass[preprint,double-spaced,showpacs,preprintnumbers,amsmath,amssymb]{revtex4}

\usepackage{graphicx}
\usepackage{dcolumn}
\usepackage{bm,slashed}

\begin{document}

\preprint{}

\title{One real function instead of the Dirac spinor function}

\author{Andrey Akhmeteli}
 \email{akhmeteli@ltasolid.com}
 \homepage{http://www.akhmeteli.org}
\affiliation{%
LTASolid Inc.\\
10616 Meadowglen Ln 2708\\
Houston, TX 77042, USA}

Copyright (2011) American Institute of Physics. This article may be downloaded for personal use only. Any other use requires prior permission of the author and the American Institute of Physics.
The following article appeared in the Journal of Mathematical Physics, \textbf{52}, 082303 (2011) (in a somewhat modified form) and may be found at
http://link.aip.org/link/?jmp/52/082303 and (free access for personal use) at http://akhmeteli.org/wp-content/uploads/2011/08/JMAPAQ528082303.pdf


\date{\today}

\begin{abstract}
Three out of four complex components of the Dirac spinor can be algebraically eliminated from the Dirac equation (if some linear combination of electromagnetic fields does not vanish), yielding a partial differential equation of the fourth order for the remaining complex component. This equation is generally equivalent to the Dirac equation. Furthermore, following Schr\"{o}dinger (Nature, \textbf{169}, 538 (1952)), the remaining component can be made real by a gauge transform, thus extending to the Dirac field the Schr\"{o}dinger's conclusion that charged fields do not necessarily require complex representation. One of the two resulting real equations for the real function describes current conservation and can be obtained from the Maxwell equations in spinor electrodynamics (the Dirac-Maxwell electrodynamics). As the Dirac equation is one of the most fundamental, these results both belong in textbooks and can be used for development of new efficient methods and algorithms of quantum chemistry.
\end{abstract}

\maketitle

\section{\label{sec:level1}Introduction}

The Dirac equation is a most fundamental equation, crucial for high energy physics and significant even for such mundane devices as lead-acid batteries (Ref.~\cite{Ahuja}). It is well-known that two components of the Dirac spinor can be algebraically eliminated from the Dirac equation (Ref.~\cite{Fay}, p.445). However, to the best of this author's knowledge, it was not noticed before that another component can be algebraically eliminated from the resulting equations, yielding a partial differential equation of the fourth order for the remaining component. While it was noted in Ref.~\cite{Bagrov} (p.28) that "the set of four first-order equations comprising the Dirac equation is generally equivalent to, and may be reduced to, a single fourth order equation", according to the authors of  Ref.~\cite{Bagrov}, this remark relates to currently known rigorous solutions of the Dirac equation, and this result had not been established for the Dirac equation with arbitrary given electromagnetic field (V.G. Bagrov, private communication).

It is possible to make one step further and rewrite the Dirac equation in terms of just one real function, rather than a complex function, using Schr\"{o}dinger's approach, who noted  (Ref.~\cite{Schroed}) that for each solution of the equations of scalar electrodynamics (the Klein-Gordon-Maxwell electrodynamics) there is a physically equivalent (i.e. coinciding with it up to a gauge transform) solution with a real matter field, despite the widespread belief about charged fields requiring complex representation. Schr\"{o}dinger concludes his work (Ref.~\cite{Schroed}) with the following: "One is interested in what happens when [the Klein-Gordon equation] is replaced by Dirac's wave equation of 1927, or other first-order equations. This and the bearing on Dirac's 1951 theory will be discussed more fully elsewhere." To the best of this author's knowledge, Schr\"{o}dinger did not publish any sequel to Ref.~\cite{Schroed}, and the lack of extension to the Dirac equation may explain the fact that Schr\"{o}dinger's work did not get the attention it deserves. Such an extension is proposed here, as the only complex component of the fourth order partial differential equation equivalent to the Dirac equation can be made real in the same way. Thus, the Dirac equation can be rewritten as an equation for one real function, rather than for a Dirac spinor. This result is both important on its own and can be used for development of new efficient methods and algorithms of quantum chemistry.
\maketitle

\section{\label{sec:level1}Elimination of components from the Dirac equation}

To illustrate the parallels with beautiful but little-known Schr\"{o}dinger's work ~\cite{Schroed}, the Dirac equation is considered as part of spinor electrodynamics, although most derivations are valid without any changes for the Dirac equation in electromagnetic field independently of the Maxwell equations.

Schr\"{o}dinger considered the equations of scalar electrodynamics:
\begin{equation}\label{eq:pr7}
(\partial^\mu+ieA^\mu)(\partial_\mu+ieA_\mu)\psi+m^2\psi=0,
\end{equation}
\begin{equation}\label{eq:pr8}
\Box A_\mu-A^\nu_{,\nu\mu}=j_\mu,
\end{equation}
\begin{equation}\label{eq:pr9}
j_\mu=ie(\psi^*\psi_{,\mu}-\psi^*_{,\mu}\psi)-2e^2 A_\mu\psi^*\psi.
\end{equation}
He observed  that  the complex charged matter field $\psi$ can be made real by a gauge transform (at least locally), although it is generally believed that complex functions are required to describe charged fields.

The equations of motion in the relevant gauge (unitary gauge (Refs.~\cite{Deumens2},\cite{Itzykson})) for the transformed 4-potential of electromagnetic field $B^{\mu}$ and real matter field $\varphi$ are as follows:
\begin{equation}\label{eq:pr10}
\Box\varphi-(e^2 B^\mu B_\mu-m^2)\varphi=0,
\end{equation}
\begin{equation}\label{eq:pr11}
\Box B_\mu-B^\nu_{,\nu\mu}=j_\mu,
\end{equation}
\begin{equation}\label{eq:pr12}
j_\mu=-2e^2 B_\mu\varphi^2.
\end{equation}

Work ~\cite{Schroed} has another unique feature. While the initial Klein-Gordon equation~(\ref{eq:pr7}) actually contains two equations for the real and imaginary components of the complex  field $\psi$, the relevant equation (\ref{eq:pr10}) for the real field $\varphi$ contains just one equation. What happened to the other equation? The missing equation is equivalent to the current conservation equation, and the latter can be derived from the Maxwell equations (\ref{eq:pr11}), as the divergence of the antisymmetric tensor vanishes.

It turns out, however, that Schr\"{o}dinger's results also hold in the case of spinor electrodynamics. In general, four complex components of the Dirac spinor function cannot be made real by a single gauge transform, but three complex components out of four can be eliminated from the Dirac equation in a general case, yielding a fourth-order partial differential equation for the remaining component, which component can be made real by a gauge transform. The resulting two equations for one real component can be replaced by one equation plus the current conservation equation, and the latter can be derived from the Maxwell equations.

Spinor electrodynamics is a more realistic theory than scalar electrodynamics, so it seems important that the charged field of spinor electrodynamics can also be described by one real function. It is not clear if similar results can be obtained for the Standard Model.

The resulting system of equations for the electromagnetic field and the real matter field is overdetermined. This fact suggests that it may be feasible to eliminate the matter field altogether, as this was done in the case of scalar electrodynamics (Ref.~\cite{Akhm12}).

Let us start with the equations of (non-second-quantized) spinor electrodynamics:
\begin{equation}\label{eq:pr25}
(i\slashed{\partial}-\slashed{A})\psi=\psi,
\end{equation}
\begin{equation}\label{eq:pr26}
\Box A_\mu-A^\nu_{,\nu\mu}=e^2\bar{\psi}\gamma_\mu\psi,
\end{equation}
where, e.g., $\slashed{A}=A_\mu\gamma^\mu$ (the Feynman slash notation). For the sake of simplicity, a system of units is used where $\hbar=c=m=1$, and the electric charge $e$ is included in $A_\mu$ ($eA_\mu \rightarrow A_\mu$).
In the chiral representation of $\gamma$-matrices (Ref.~\cite{Itzykson})
\begin{equation}\label{eq:d1}
\gamma^0=\left( \begin{array}{cc}
0 & -I\\
-I & 0 \end{array} \right),\gamma^i=\left( \begin{array}{cc}
0 & \sigma^i \\
-\sigma^i & 0 \end{array} \right),
\end{equation}
where index $i$ runs from 1 to 3, and $\sigma^i$ are the Pauli matrices.
If $\psi$ has components
\begin{equation}\label{eq:d2}
\psi=\left( \begin{array}{c}
\psi_1\\
\psi_2\\
\psi_3\\
\psi_4\end{array}\right),
\end{equation}
the Dirac equation (\ref{eq:pr25}) can be written in components as follows:
\begin{eqnarray}\label{eq:d3}
\nonumber
(A^0+A^3)\psi_3+(A^1-\imath A^2)\psi_4+\\
+\imath(\psi_{3,3}-\imath\psi_{4,2}+\psi_{4,1}-\psi_{3,0})=\psi_1,
\end{eqnarray}
\begin{eqnarray}\label{eq:d4}
\nonumber
(A^1+\imath A^2)\psi_3+(A^0-A^3)\psi_4-\\
-\imath(\psi_{4,3}-\imath\psi_{3,2}-\psi_{3,1}+\psi_{4,0})=\psi_2,
\end{eqnarray}
\begin{eqnarray}\label{eq:d5}
\nonumber
(A^0-A^3)\psi_1-(A^1-\imath A^2)\psi_2-\\
-\imath(\psi_{1,3}-\imath\psi_{2,2}+\psi_{2,1}+\psi_{1,0})=\psi_3,
\end{eqnarray}
\begin{eqnarray}\label{eq:d6}
\nonumber
-(A^1+\imath A^2)\psi_1+(A^0+A^3)\psi_2+\\
+\imath\psi_{2,3}+\psi_{1,2}-\imath(\psi_{1,1}+\psi_{2,0})=\psi_4.
\end{eqnarray}
Obviously, equations (\ref{eq:d5},\ref{eq:d6}) can be used to express components $\psi_3,\psi_4$ via $\psi_1,\psi_2$ and eliminate them from equations (\ref{eq:d3},\ref{eq:d4}) (cf. Ref.~\cite{Fay}, p.445). The resulting equations for $\psi_1$ and $\psi_2$ are as follows:
\begin{widetext}
\begin{eqnarray}\label{eq:d7}
\nonumber
-\psi_{1,\mu}^{,\mu}+\psi_2(-\imath A^1_{,3}-A^2_{,3}+A^0_{,2}+A^3_{,2}+
\imath(A^0_{,1}+A^3_{,1}+A^1_{,0})+A^2_{,0})+\\
+\psi_1(-1+A^{\mu} A_{\mu}-\imath A^{\mu}_{,\mu}+\imath A^0_{,3}-A^1_{,2}+A^2_{,1}+\imath A^3_{,0})-2\imath A^{\mu}\psi_{1,\mu}=0,
\end{eqnarray}
\begin{eqnarray}\label{eq:d8}
\nonumber
-\psi_{2,\mu}^{,\mu}+\imath\psi_1( A^1_{,3}+\imath A^2_{,3}+\imath A^0_{,2}-\imath A^3_{,2}+
A^0_{,1}-A^3_{,1}+A^1_{,0}+\imath A^2_{,0})+\\
+\psi_2(-1+A^{\mu} A_{\mu}-\imath( A^{\mu}_{,\mu}+A^0_{,3}+\imath A^1_{,2}-\imath A^2_{,1}+ A^3_{,0}))-2\imath A^{\mu}\psi_{2,\mu}=0.
\end{eqnarray}
\end{widetext}

As equation (\ref{eq:d7}) contains $\psi_2$, but not its derivatives, it can be used to express $\psi_2$ via $\psi_1$:
\begin{eqnarray}\label{eq:d8nn1}
\psi_2=-\left(\imath F^1+F^2\right)^{-1}\left(\Box'+\imath F^3\right)\psi_1,
\end{eqnarray}
where $F^i=E^i+\imath H^i$, electric field $E^i$ and magnetic field $H^i$ are defined by the standard formulae
\begin{eqnarray}\label{eq:d8n1}
F^{\mu\nu}=A^{\nu,\mu}-A^{\mu,\nu}=\left( \begin{array}{cccc}
0 & -E^1 & -E^2 & -E^3\\
E^1 & 0 & -H^3 & H^2\\
E^2 & H^3 & 0 & -H^1\\
E^3 &-H^2 & H^1 & 0  \end{array} \right),
\end{eqnarray}
and the modified d'Alembertian $\Box'$ is defined as follows:
\begin{eqnarray}\label{eq:d8n2}
\Box'=\partial^\mu\partial_\mu+2\imath A^\mu\partial_\mu+\imath A^\mu_{,\mu}-A^\mu A_\mu+1.
\end{eqnarray}
Using the above notation, equation (\ref{eq:d8}) can be rewritten as follows:
\begin{eqnarray}\label{eq:d8nn2}
-\left(\Box'-\imath F^3\right)\psi_2-\left(\imath F^1-F^2\right)\psi_1=0,
\end{eqnarray}
so equation(\ref{eq:d8nn1}) can be used to eliminate $\psi_2$ from equation (\ref{eq:d8nn2}), yielding an equation of the fourth order for $\psi_1$:
\begin{eqnarray}\label{eq:d8n1}
\left(\left(\Box'-\imath F^3\right)\left(\imath F^1+F^2\right)^{-1}\left(\Box'+\imath F^3\right)-\imath F^1+F^2\right)\psi_1=0.
\end{eqnarray}

This equation is equivalent to the Dirac equation (if $\imath F^1+F^2\slashed{\equiv}0$).

It should be noted that the coefficient at $\psi_2$ in equation (\ref{eq:d7}) is gauge-invariant (it can be expressed via electromagnetic fields). While this elimination could not be performed for zero electromagnetic fields, this does not look like a serious limitation, as in reality there always exist electromagnetic fields in the presence of charged fields, although they may be very small. However, it should be noted that free spinor field presents a special case and is not considered in this work, as it does not satisfy the equations of spinor electrodynamics. It is not clear how free field being a special case is related to the divergencies in quantum electrodynamics. It should also be noted that the above procedure could be applied to any component of the spinor function, not just to $\psi_1$, yielding equations similar to equation (\ref{eq:d8n1}). Presenting the above results in a more symmetric form is beyond the scope of this work.

While the above elimination of the third component of the Dirac spinor is straightforward, this author failed to find it elsewhere, but cannot be sure that this important result, which belongs in textbooks, was not published previously.

Using a gauge transform, it is possible to make $\psi_1$ real (at least locally). Then the real and the imaginary parts of equation (\ref{eq:d8}) after substitution of the expression for $\psi_2$ will present two equations for $\psi_1$. However, it is possible to construct just one equation for $\psi_1$ in such a way that the system containing this equation and the current conservation equation will be equivalent to equation (\ref{eq:d8}).

Let us consider the current conservation equation:
\begin{equation}\label{eq:d9}
(\bar{\psi}\gamma^\mu\psi)_{,\mu}=0,
\end{equation}
or
\begin{equation}\label{eq:d10}
(\bar{\psi}_{,\mu}\gamma^\mu\psi)+(\bar{\psi}\gamma^\mu\psi_{,\mu})=0,
\end{equation}
On the other hand,
\begin{eqnarray}\label{eq:d11}
\nonumber
(\bar{\psi}_{,\mu}\gamma^\mu\psi)^*=(\bar{\psi}_{,\mu}\gamma^\mu\psi)^\dag=\\
=\psi^\dag(\gamma^\mu)^\dag(\psi^\dag_{,\mu}\gamma^0)^\dag=
\bar{\psi}\gamma^0\gamma^0\gamma^\mu\gamma^0\gamma^0\psi_{,\mu}=\bar{\psi}\gamma^\mu\psi_{,\mu},
\end{eqnarray}
as $(\gamma^\mu)^\dag=\gamma^0\gamma^\mu\gamma^0$ (Ref.~\cite{Itzykson}), so the current conservation equation can be written as follows:
\begin{eqnarray}\label{eq:d12}
2\textrm{Re}(\bar{\psi}\gamma^\mu\psi_{,\mu})=2\textrm{Im}(\bar{\psi}\imath\slashed{\partial}\psi)=\\
\nonumber
=2\textrm{Im}(\bar{\psi}(\imath\slashed{\partial}-\slashed{A}-1)\psi)=0,
\end{eqnarray}
as it is not difficult to check that values $\bar{\psi}\slashed{A}\psi$ and $\bar{\psi}\psi$ are real.

If equations (\ref{eq:d5},\ref{eq:d6},\ref{eq:d7}) hold, only the second component of spinor $(\imath\slashed{\partial}-\slashed{A}-1)\psi$ can be nonzero, so equation (\ref{eq:d12}) can be written as follows:
\begin{equation}\label{eq:d13}
2\textrm{Im}(-\psi_4^* \delta)=0,
\end{equation}
where $\delta$ is the left-hand side of equation (\ref{eq:d8}). Therefore, if $\psi_4$ does not vanish identically, the system containing the current conservation equation equation (\ref{eq:d13}) and the following equation
\begin{equation}\label{eq:d14}
2\textrm{Re}(\psi_4^* \delta)=0,
\end{equation}
is equivalent to equation (\ref{eq:d8}).

\maketitle

\section{\label{sec:level1}Conclusion}

Thus, three complex components of the Dirac equation out of four can be algebraically eliminated, and the remaining component can be made real using a gauge transform. In particular, most results of work ~\cite{Schroed} for scalar electrodynamics are extended to a more realistic theory -- spinor electrodynamics: in a certain gauge, the charged matter field can be described by one real function. This function satisfies a real partial differential equation (of the fourth order) and the current conservation equation, which can be obtained from the Maxwell equations. Thus, the system of equations for the function is overdetermined. As it was shown in work ~\cite{Akhm12} that the matter field can be naturally eliminated from the equations of scalar electrodynamics, the results of this work suggest that the spinor field can be eliminated from spinor electrodynamics as well (and therefore spinor electrodynamics can embedded into a quantum field theory in the same way as scalar electrodynamics (cf. Ref.~\cite{Akhm12}) , but this has not been proven yet.

\section*{Acknowledgments}

The author is grateful to V.G. Bagrov, A. Yu. Kamenshchik, and A. E. Allahverdyan for their interest in this work and valuable remarks and to J. Noldus for useful discussions.

.

\end{document}